Title:

Ecuador's mangrove forest carbon stocks: A spatiotemporal analysis of living carbon holdings and their depletion since the advent of commercial aquaculture

Authors and Affiliations:

Corresponding Author:
Stuart E. Hamilton, Ph.D
Assistant Professor
Department of Geography and Geosciences
Salisbury University
Salisbury, MD 21801
410-543-6460
stuartehamilton@gmail.com

Other Author:
John Lovette
Ph.D Candidate
Dept. Geography
University of North Carolina at Chapel Hill
Chapel Hill, NC 27599
919-962-1558
jplovette@gmail.com

**ABSTRACT**

In this paper we estimate the living carbon lost from Ecuador's mangrove forests since the advent of export-focused shrimp aquaculture. We use remote sensing techniques to delineate the extent of mangroves and aquaculture at approximately decadal periods since the arrival of aquaculture in each Ecuadorian estuary. We then spatiotemporally calculate the carbon values of the mangrove forests and estimate the amount of carbon lost due to direct displacement by aquaculture. Additionally, we calculate the new carbon stocks generated due to mangrove reforestation or afforestation. This research introduces time and land use / land cover change (LUCC) into the tropical forest carbon literature and examines forest carbon loss at a higher spatiotemporal resolution than in many earlier analyses. We find that 80%, or 7,014,517 t of the living carbon lost in Ecuadorian mangrove forests can be attributed to direct displacement of mangrove forests by shrimp aquaculture. We also find that Intergovernmental Panel on Climate Change (IPCC) compliant carbon grids within Ecuador's estuaries overestimate living carbon levels in estuaries where substantial LUCC has occurred. By approaching the mangrove forest carbon loss question from a LUCC perspective, these findings allow for tropical nations and other intervention agents to prioritize and target a limited set of land transitions that likely drive the majority of carbon losses. This singular cause of transition has implications for programs that attempt to offset or limit future forest carbon losses and place value on forest carbon or other forest good and services.

**INTRODUCTION**

Tropical deforestation is the second largest cause of global greenhouse gas emissions behind burning of fossil fuels and is responsible for releasing on average 1.4 Pg C yr$^{-1}$ between 1980 and 2005 (Baccini et al., 2012; Defries et al., 2002; Houghton, 2003; Melillo, Houghton, Kicklighter, & McGuire, 1996). Tropical forests contain the highest carbon reservoirs of all global forests with between 228.7 Pg C (Baccini et al., 2012) and 247 Pg C (Saatchi et al., 2011) stored within them. This equates to 55% of global forest carbon (Pan et al., 2011). It has been suggested that these global estimates of tropical forest carbon stocks, and similarly those of emissions, are likely underestimations due to the fact that the current levels of carbon stored in tropical mangroves and other organic-rich peatlands, particularly belowground, remain relatively unknown and unaccounted for in many global analyses (Pan et al., 2011; Donato et al., 2011; Ladd et al., 2013; Zarin, 2012).

It has been estimated that global mangrove forests contain between 937 t C ha$^{-1}$ and 1023 t C ha$^{-1}$ (Donato et al., 2011; Alongi, 2012) with higher biomass, and hence higher carbon densities closer to the equator (Saenger & Snedaker, 1993; Twilley, Chen, & Hargis, 1992). This calculation of mangrove forest carbon storage per unit area is approximately three to four times higher than that of other tropical forests types that only average between 223 t C ha$^{-1}$ and 316 t C ha$^{-1}$ (IPCC, 2001). For this reason, mangrove deforestation has the potential to release more $CO_2$ per unit area that almost any other global forest type. Recent work on carbon within mangrove forests, both aboveground and belowground, is expanding and is even placing economic values on these potential carbon reservoirs. For example, in addition to the recent creation of one time snapshots of whole-system carbon levels in mangrove forests (Donato et al., 2011) others have attempted to apply an economic value to mangrove carbon sinks (Siikamäki,

Sanchirico, & Jardine, 2012). Although such snapshot mangrove carbon storage studies are spatial in nature, few spatiotemporal carbon-based analyses of mangroves appear to exist and even fewer focus on specific land use / land cover transitions, such as mangrove to aquaculture conversion, that are likely responsible for the majority of the carbon losses.

We use a unique high-resolution 10 m by 10 m LUCC grid spread across the majority of Ecuador's estuaries to determine mangrove carbon holdings and account for factors driving mangrove biomass such as mangrove latitude (Saenger & Snedaker, 1993; Twilley et al., 1992), mangrove intra-estuarine location (Chen & Twilley, 1999; Sherman, Fahey, & Martinez, 2003), and mangrove species type (Sherman et al., 2003; Komiyama, Ong, & Poungparn, 2008). In doing so we not only present estimates of current and historic mangrove carbon levels, but more importantly we document the actual land use / land cover transitions that are responsible for the majority of carbon losses over the analysis period.

The 1980s and 1990s growth of aquaculture is well documented (Naylor et al., 1998; Naylor et al., 2000) and shows no sign of abating (Figure 1). As of 2012 seafood production via aquaculture almost outstripped that of wild catch, with production levels of 90.43 and 91.3 million t respectively (FAO 2014). With fisheries capture production declining and aquaculture production expanding it is likely that aquaculture has already passed capture as the primary source of global seafood production. Within Ecuador the expansion of aquaculture exceeds the global trend (Figure 1b). From essentially nothing in the early 1980s, shrimp aquaculture has grown to a $1.39 billion industry by 2012 and is now the second largest component of the Ecuadorian economy after fossil fuels. This expansion is almost entirely attributable to shrimp aquaculture (Figure 1b) and has led to land use / land cover transitions in Ecuadorian estuaries

with both historic mangrove and other estuarine land cover now converted to shrimp ponds (Hamilton & Stankwitz 2012).

Figure 1a

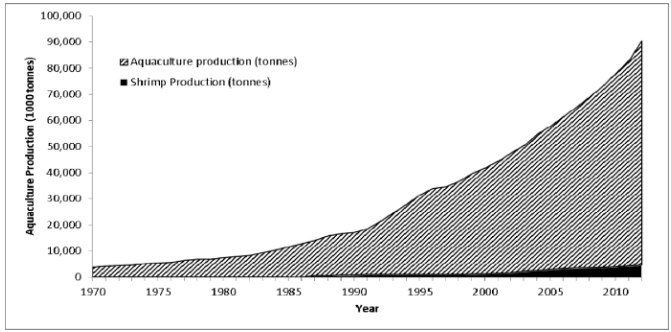

Figure 1b

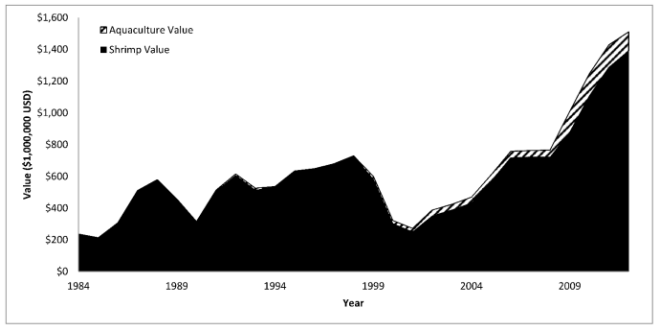

Figure 1. Global and Ecuadorian aquaculture and shrimp aquaculture growth (FAO 2014). Figure 1a depicts the global growth in aquaculture from a nominal amount in 1970 to greater than 90 million t in 2012. Figure 1b depicts the growth rate of shrimp aquaculture in Ecuador from approximately 200 million USD in 1984 to approximately 1.4 billion USD in 2012.

It is well established that non-mangrove tropical forests are often converted to agriculture, resulting in increased levels of atmospheric carbon (Pan et al., 2011; Geist & Lambin, 2002). On the other hand, mangrove forests globally are most at risk from conversion to aquaculture, as opposed to agriculture, with estimates as high as 28-40% of the total global mangrove area being already converted to aquaculture (Hamilton, 2013; Polidoro et al., 2010).

Of all regions with mangrove to aquaculture conversion, coastal Ecuador has likely undergone the highest levels of transformation with approximately 40% of mangrove converted to aquaculture with certain regions experiencing almost total mangrove to aquaculture conversion (Hamilton, 2013). Despite mangrove to aquaculture conversion rates that far outstrip other tropical forest to agriculture conversion rates, and the fact that mangrove forests having a far higher carbon level per unit area than other tropical forests, a paucity of research exists examining the mangrove conversion question in terms of changes in carbon stocks over space and time.

**STUDY AREA**

The study area consists of all the major coastal estuaries of mainland Ecuador, with the exception of the mouth of the Guayas River near the city of Guayaquil and the Galapagos Islands. For security reasons, the Instituto Geográfico Militar does not release historical aerial photographs of this portion of Guayas province and thus it has been excluded from this study. The Galapagos is excluded due to its remote location away from the Ecuadorian mainland. Ecuador was selected for analysis due to its long history of estuarine shrimp aquaculture, availability of high-resolution spatiotemporal data for each estuary, pre-established mangrove and aquaculture surveys, participation in payment for performance carbon programs, and tropical location on the equator. The combined area of our study area is 201,151 1 ha grid cells across all estuaries resulting in 201,151 x $10^2$ 10 m by 10 m LUCC analysis cells.

From north to south, the Ecuadorian estuaries analyzed (Figure 2 & Supplemental Map Package) are: (i) Cayapas-Mataje, located wholly within Esmeraldas province along the Colombian border in and around the town of San Lorenzo; (ii) Muisné, located wholly within Esmeraldas province near the town of the same name; (iii) Cojimíes, located on the border between Esmeraldas and Manabí provinces in and around the city of Pedernales; (iv) Chone estuary, located wholly within Manabí province in and around the city of Bahia de Caráquez; (v) Isla Puná; an island in the Gulf of Guayaquil, (vi) the entire coastal region of El Oro province in and around the city of Machala from the southern edge of Guayas province in the north to the major estuary known as Grande Estuary on the Peruvian border in the south. We estimate these regions to comprise greater than 95% of the historic mangrove habitat in Esmeraldas, Manabí, and El Oro provinces and approximately 26% of the historic pre-aquaculture forest in Guayas province.

Figure 2. Study Sites.

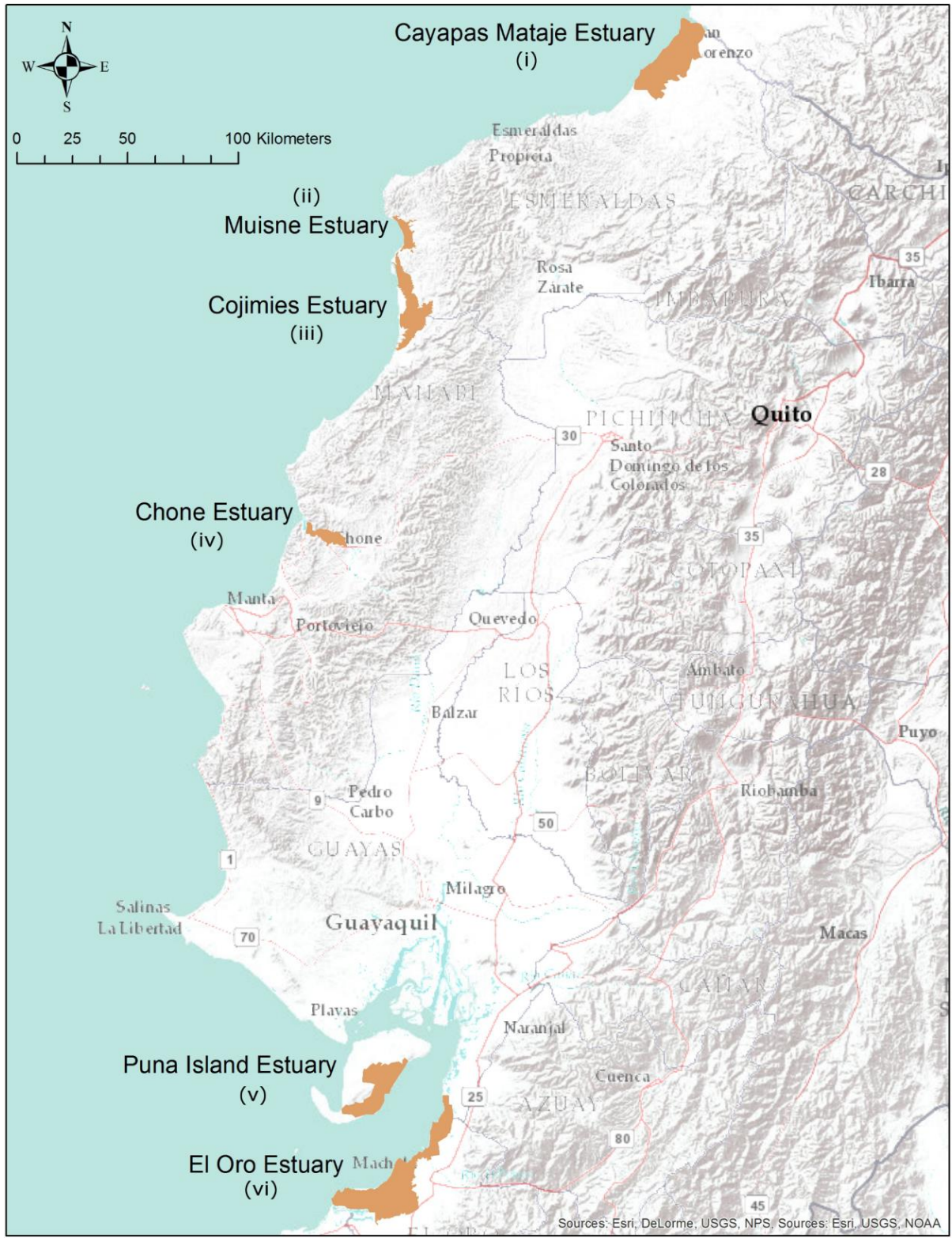

As of 2013, all remaining mangrove stands in Ecuador are protected at the federal level. Prior to the national protection decree issued in 2013, the mangroves in each study area had varying levels of protected status beginning at different times. The mangroves of Cayapas-Mataje (i) are almost wholly contained within an original RAMSAR site with the federal government as the long-term legal owner of the estuary and the Ministry of the Environment overseeing the mangrove resource within the estuary since at least 1995. Since 2003 approximately 1% of the mangroves in Muisné (ii) estuary are privately protected as the Muisné River Estuary Wildlife Reserve while the majority of the estuary appears without protection. According to the Ministry of the Environment SNAP (National System of Protected Areas) database, conversations with local fisherman, and a literature search; Cojimíes (iii) appears to have no government support or protected status at any level. Since 2002, Chone estuary (iv) has a small portion of the estuary protected as the Corazón and Frigatas Islands Wildlife Reserve (Registro Oficial No 733). Chone estuary in its entirety has been covered under a voluntary special area management since 1988 with the goal of improving the health of the estuary and surrounding area (Arriaga, Montaño, & Vásconez, 1999). The mangroves of the Gulf of Guayaquil and the Guayas River Estuary within Guayas Province (v) have federal protection along the eastern portion of the estuary but this is outside of our analysis area and none of the other areas analyzed have protection beyond the recent national decree. This lack of historic protected status extends to Isla Puná within the Gulf of Guayaquil. El Oro (vi) has no protection beyond the recent national decree, although federal protection within the estuary exists once you cross the international border into northern Peru.

**MATERIALS AND METHODS**

Field research permission for ground truthing in Manabí province was obtained from MAE (Ministry of the Environment) Manabí office in Portoviejo and field permission for ground truthing in Esmeraldas province was obtained from the MAE offices in San Lorenzo and Muisné Ecuador. Ground truthing was conducted by the authors of this paper in combination with local staff from the Ministry of the Environment.

*Depicting Land Use Cover Change*

Each of the 6 study area estuaries (Figure 2) was divided into 10 m x 10 m LUCC analysis grids. We overlaid the grids on each of the estuaries which themselves were delineated from 1:25,000 scale topographic maps. Each of these 100 $m^2$ LUCC grids were then aggregated into 1 ha carbon grids. To obtain the initial survey data, the Landsat archive at the Global Land Cover facility was queried to determine the first appearance of aquaculture in each estuary. Once this was ascertained, the first usable Landsat image previous to this date was obtained. For all estuaries aside from Puná and Cayapas-Mataje, the earliest Landsat images clearly had aquaculture. For these estuaries, a combination of air photos and 1:25,000 topographic maps were utilized to depict LUCC at the initial survey date. During the period of Landsat 7 line scan problems in the late 2000s, we utilized ASTER to compliment Landsat datasets. The final survey used Rapid Eye imagery to supplement the coarser remote sensing data. These utilized instruments have varying spatial resolutions between 1 m and 30 m. Landsat at 30 m and Aster at 15 m are coarser than the 10 m grid resolution, leading to duplication of values across neighboring cells; this was overcome by aggregating data into the 1 ha carbon analyses grids for reporting. At this reporting unit even the coarsest dataset will have 1089 inputs into each grid.

No spectral method exists for detection of aquaculture, therefore shrimp ponds were digitized manually whereas the mangrove areas were extracted via the standard process of IsoData-driven unsupervised clusters that were then identified using ancillary data and field observations. We then converted the IsoData-derived pixels into polygons via a process of manual digitization assisted by an NDVI layer created for each estuary. Ground-truthing is not possible for earlier surveys but was conducted in 2014 for the 2011 survey in all estuaries across all LC (land-cover) classes.

Upon completion of the LC analysis for each time period in each estuary, each of the sub-grids was coded with a RS derived LC value of either mangrove, aquaculture, other aquatic or other terrestrial, depicted by a binary value of 0, indicating absence, or a value of 1, indicating majority presence. From these 100 sub-grid binary values per 1 ha grid, the 1 ha cells were coded with a continuous value that represents the percentage of the cell that is mangrove, aquaculture, or other (terrestrial and aquatic) from a minimum of 0% to a maximum of 100%. This LC type is a grid level variable represented in the data as a continuous value from 0 to 1, and expressed as $M_{it}$ (mangrove density at location $i$ and time period $t$), $Aq_{it}$ (aquaculture density at location $i$ and time period $t$), or $O_{it}$ (other at location $i$ and time period $t$), which is either surface water with limited mud flats and salt pans or non-mangrove and non-aquaculture terrestrial environments. For example, if 50 of the 10 m sub-grids in any 1 ha cell during any survey period are mangrove, the value for $M_{it}$ in the 1 ha cell would be 0.5. The LC values in each 1 ha cell sum to 1. A maximum theoretical of $1.01 \times 10^{14}$ combination of $LC_{it}$ combinations exist at the 1 ha grid level with $1.01 \times 10^{16}$ possible $LC_{it}$ combination inputs at the 10 m sub-grid level and $1.01 \times 10^{22}$ when considering different mangrove species.

*Estimating Mangrove Carbon*

Across all forest types, including mangrove, biomass is utilized as a proxy for living carbon storage. Mangrove biomass across all species is proportional to the ambient isolation or solar energy at each mangrove location, therefore latitude can be used to account for most of the variability of biomass within a mangrove forest at differing locations (Saenger & Snedaker, 1993; Twilley et al., 1992). This fact likely explains why the tallest *Rhizophora* mangroves in the world are found straddling the equator within northern Ecuador. Upon completion of the remote sensing analysis mangrove living carbon estimates were generated via the four methods listed below.

Synthesizing the peer-reviewed above-ground mangrove biomass (AGMB) estimates across 11 nations and 5 dominant species, a 1993 study utilized linear regression to calculate biomass as a function of latitude and reported that 69% of the variance in aboveground mangrove biomass (AGMB) can be explained solely by latitude (Saenger & Snedaker, 1993). Using this linear model, AGMB was calculated as a function of latitude across all estuaries, in all grid cells, and at all time periods. Once AGMB grids were created, belowground mangrove biomass (BGMB) was calculated as an allometrically derived function of AGMB across all estuaries, in all grid cells, and during all time periods.

Synthesizing data from global mangrove studies across the peer-reviewed literature, the AGMB : BGMB ratio is shown to average *1 : 0.52* (DeFries et al., 2002). Combined mangrove biomass (CMB) can thus be expressed simply as CMB = BGMB + AGMB, but using the AGMB : BGMB conversion factors, CMB can also be expressed as CMB = (1 + AGMB : BGMB) * AGMB. Using the methods defined by Saenger & Snedaker (1993) and DeFries et al. (2002), CMB was calculated across all estuaries, in all grid cells $i$, and during all time periods $t$ and then converted into combined aboveground and belowground carbon (CC). This conversion from CMD to C$C^*$ was conducted using the mangrove biomass to carbon ratio of *1 : 0.464* (Donato et al., 2011; Kauffman, Header, Cole, Dire, & Donato, 2011). The .464 value is approximately constant in the literature with other values expressed between .45-.50 (Twilley et al., 1992; Kauffman & Donato, 2012). All equations are shown in the format: CC = $M_{it}$ * (Biomass : Carbon) * CMB. Equation (1) depicts the combined function.

Equation (1)
$$CC^*_i\,(t.ha^{-1}) = (M_{it}\,(.464\,(373.273 - 8.486\,|Lat|)))$$

$CC^*$ = combined carbon, $t$ = tonnes, $ha$ = hectare, $M_{it}$ = Mangrove density at grid location $i$ and time slice $t$ on a scale of 0 to 1, $|Lat|$ = absolute latitude.

Other studies also address the mangrove biomass question from a latitude perspective using a linear modeling (Twilley et al., 1992). This research reports that 75% of the variance in AGMB within mangroves can be explained solely by latitude (Equation 2). By using an AGMB : BGMB ratio of *1: 0.82* CMB was calculated across all estuaries, in all grid cells *i*, and during all time periods *t* and then converted into CC using the .464 value from Equation 1.

Equation (2)
$$CC_i\,(t.ha^{-1}) = (M_{it}\,(.464\,(543.27 - 13.269\,|Lat|)))$$

$CC$ = combined carbon, $t$ = tonnes, $ha$ = hectare, $M_{it}$ = Mangrove density at grid location $i$ and time slice $t$ on a scale of 0 to 1, $|Lat|$ = absolute latitude.

The mangrove ecosystem along the Ecuadorian coast is dominated by three species: *Rhizophora mangle*, *Laguncularia racemosa*, and *Avicennia germinans*. Aside from general species availability in the region, a variety of geophysical factors influence the species' distributions in each estuary, including but not limited to: soil and water salinity, nutrient availability, tidal dynamics, wave-action tolerance, and geomorphologic processes. Five zonation classes have been traditionally used to characterize entire mangrove systems (Lugo & Snedaker, 1974; Pool, Snedaker, & Lugo, 1977). However, we move past these generic classes. Using spatial proxies for salinity, tidal inundation, and wave-action tolerance, a species likelihood model was constructed within the grids for each of the six studied estuaries.

First, using a priori knowledge of mangrove species distribution in the estuaries from previous studies and those published in the literature, we also utilize a set species likelihood for each of the three mangrove species (Arriaga et al., 1999; Madsen et al., 2001; Spalding, Kainuma, & Collins, 2010). Red mangrove dominates the coast, therefore $M_{Rit}$ was set at 0.90. White and black mangroves make up the majority of the remainder of the mangrove stands and thus both given $M_{(s)it}$ weights of 0.05 (Equation 3).

Equation (3)

$$CC'_i (t.ha^{-1}) = (M_{it} (.464 (0.90 * BM_R + 0.05 * BM_B + 0.05 * BM_W)))$$

$CC'$ = combined carbon, $t$ = tonnes, $ha$ = hectare, $M_{it}$ = Mangrove density at grid location $i$ and time slice $t$ on a scale of 0 to 1, $BM_{(s)}$ = combined biomass of species ($s$), $_R$ = *Rhizophora mangle*, $_B$ = *Avicennia germinans*, $_W$ = *Laguncularia racemosa*

Three distance measurements were used as estimations for biogeographic factors of species distribution: distance to landward edge (salinity and freshwater input), distance to water (tidal inundation), and distance to ocean (wave action tolerance). Each of these measurements was carried out by extracting appropriate information from the land cover database and utilizing simple estuarine level distance analyses. The landward edge was defined as the furthest inland extent on all sides of the estuary that was not in contact with the ocean. The water layer was defined as all rivers and channels digitized from 5 m resolution data. The distance to ocean measurement was determined using the inlets to the estuary.

Using the three distance parameters, the likelihood of each of the three species was determined in each grid cell in each time period based on the tolerance of that species for the environmental parameter. Wave action drives a significant amount of species distribution at the water-edge of an estuary. Because of the extensive network of prop roots, the *Rhizophora mangle* shields species that are not as well suited for the immediate coastal or streamside habitat whereas *Laguncularia racemosa* and *Avicennia germinans* exist at higher elevations within the estuary, thus separated from immediate wave-action and frequent inundation (Pool et al., 1977; Sherman et al., 2003; Duke, Ball, & Ellison, 1998).

Salinity also drives species distribution, furthering zonation by each species' salinity tolerance. *Rhizophora mangle* are typically the most salt-tolerant while *Laguncularia racemosa* and *Avicennia germinans* thrive in more freshwater, upstream environments (Pool et al., 1977; Sherman et al., 2003; Duke et al., 1998; Chen & Twilley, 1999). The three distance parameters were classified for each of the species and weighted to create a theoretical species distribution across each estuary. By compiling and normalizing the species likelihood layers in each estuary grid cell, the output provides values representing the potential percentage of the 1 ha grid cell covered by *Rhizophora mangle*, *Laguncularia racemosa*, and *Avicennia germinans* as well as the area without mangrove cover. These data were then combined with LC derived mangrove percentage data at the grid level. When mangrove was present the species likelihood model was applied to give a likelihood of each species type in each grid. For example, a grid cell classified as 100% mangrove from RS methods could then have a secondary likelihood classification of .9 .05 .05, meaning that the mangroves present during the survey have a 90% likelihood of being *Rhizophora mangle*, a 5% chance of being *Laguncularia racemosa*, and a 5% chance of being *Avicennia germinans*. The likelihood of each species' presence in each grid cell $i$, at time $t$ is expressed below (Equation 4).

Equation (4)

$$CC\hat{}_i (t.ha^{-1}) = (M_{it} (.464 (nM_{Rit} * BM_R + nM_{Bit} * BM_B + nM_{Wit} * BM_W)))$$

$CC\hat{}$ = combined carbon, $t$ = tonnes, $ha$ = hectare, $M_{it}$ = Mangrove density at grid location $i$ and time slice $t$ on a scale of 0 to 1, $nM_{(s)it}$ = normalized presence likelihood of species ($s$) at grid location $i$ and time slice $t$, $BM_{(s)}$ = combined biomass of species ($s$), $R$ = *Rhizophora mangle*, $B$ = *Avicennia germinans*, $W$ = *Laguncularia racemosa*.

In the largest review of mangrove allometry (Komiyama et al., 2008), biomass is seen to be highly species-specific as opposed to site-specific, therefore making existing allometric equations viable in Ecuador. Using forest structure dynamics from similar mangrove stands and their derived allometric relationships, the living biomass of the stands in this study are estimated (Komiyama et al., 2008; Fromard et al., 1998; Soares & Schaeffer-Novelli, 2005).

In summary, the first method of calculating CC is a latitude based function of AGMB (Saenger & Snedaker, 1993) combined with BGMB measures derived from field measures to obtain CMB (Komiyama et al., 2008) (Equation 1). The second method is also a direct latitude to CMB conversion function that is established in the literature (Twilley et al., 1992; Siikamäki et al., 2012) (Equation 2). The third method is species specific and used a priori knowledge of mangrove species distribution in Ecuadorian estuaries from previous studies that emphasize the dominance of *Rhizophora mangle* within Ecuador estuaries to obtain CMB (Equation 3) (Arriaga et al., 1999; Madsen et al., 2001; Spalding et al., 2010). The final method utilized is a species likelihood model constructed for this paper to obtain CMB (Equation 4). This final model relied on the principle of mangrove zonation that is well documented in the literature (Duke et al., 1998; Snedaker, 1982). These four methods all result in differing measures of CC.

Carbon change maps were created for each time period in each estuary using all four carbon methods described above. By combining the carbon change grids with the land cover classes at each survey period, cells were given a land cover conversion value indicating the type of transition responsible for the carbon losses or gains. All possible transitions were noted to account for reforestation or afforestation. Upon completion, the magnitude of carbon change and CC transitions were reported at the estuary and national level and were also extrapolated to other selected major mangrove holding nations. Finally we compare our CC findings with those of IPCC compliant data for 2000.

**RESULTS**

At the initial pre-aquaculture survey the CC stocks across all estuaries analyzed are calculated to be 18,754,752 t C ± 27% in total (Figure 3). By 2011, the CC in these pre-existing forests had diminished by 8,813,841 t C ± 27% (Table 1) to 9,940,912 t C ± 27%. This equates to a CC loss of 47% from pre-aquaculture to 2011. The majority of the CC losses occurred between 1970 and 1990 and losses appear to have stabilized by the 2000s and remain stable to the present. Losses in El Oro province, around Grande Estuary, and in Cojimíes account for most of this change with CC losses of 3,585,069 t C ± 27% and 2,218,212 t C ± 27% respectively (Figure 3). The areas of highest CC loss as a percentage of original stock are Chone and Cojimíes, with losses of 76% and 80% respectively (Figure 3). The Ramsar site and protected forests of Cayapas-Mataje lost the least of their pre-aquaculture CC holdings with only 22% of their initial CC lost (Figure 3).

Figure 3

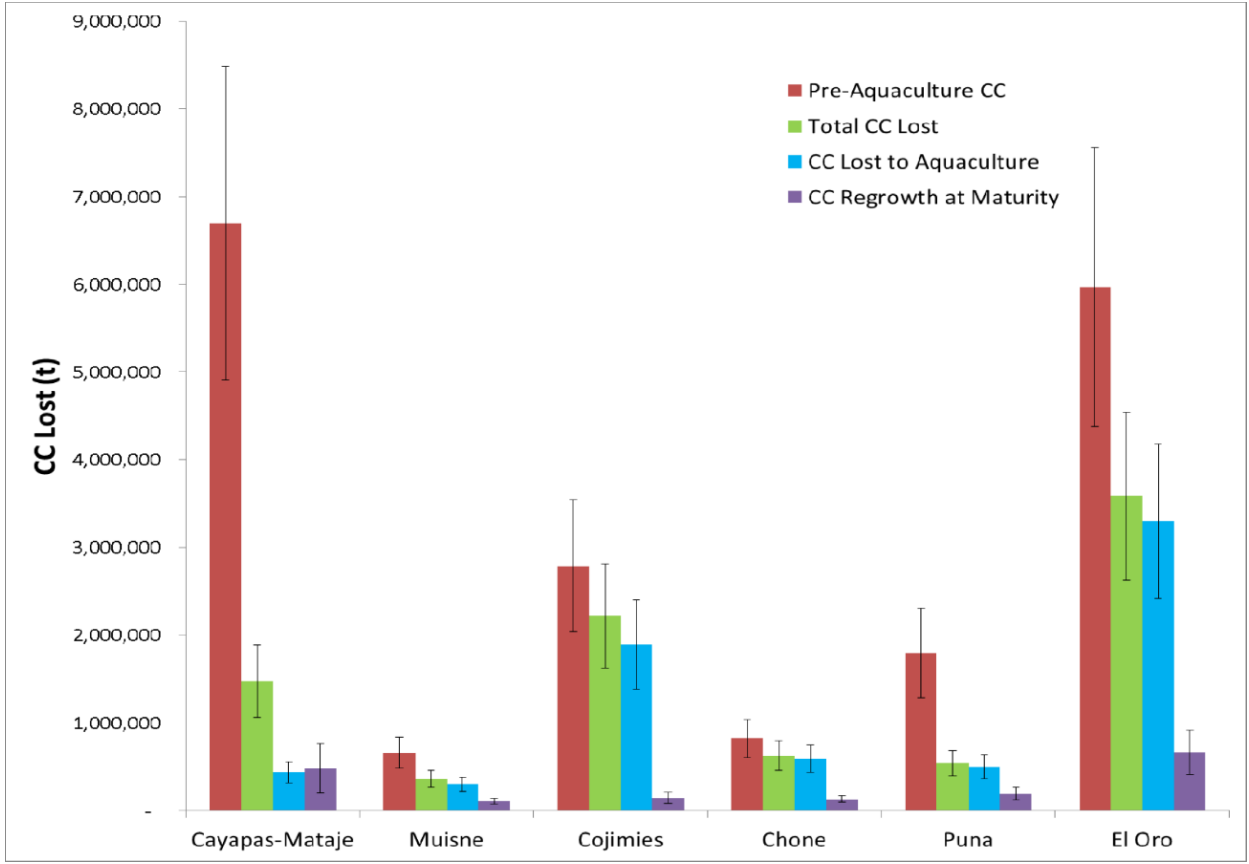

Figure 3. CC levels pre-aquaculture, CC losses from pre-aquaculture to 2011, CC lose attributable to aquaculture, and CC gains due to afforestation or reforestation). All values reported are mid-range values with the error bars representing the minimum and maximum calculated values under equations 1 – 4.

Table 1. CC losses from areas delineated as mangrove forest in the initial survey.

| Estuary | CC | CC* | CC^ | CC' | Mid | Mean |
|---|---|---|---|---|---|---|
| Cayapas-Mataje | 1887781 | 1299630 | 1059688 | 1092327 | 1473735 | 1334857 |
| Muisné | 450671 | 309815 | 263959 | 259915 | 355293 | 321090 |
| Cojimíes | 2814997 | 1935951 | 1680017 | 1621427 | 2218212 | 2013098 |
| Chone | 789448 | 542310 | 464628 | 455711 | 622580 | 563024 |
| Isla Puna | 688025 | 475251 | 395717 | 397241 | 541871 | 489059 |
| El Oro | 4541515 | 3142489 | 2713867 | 2628623 | 3585069 | 3256624 |
| **TOTAL** | **11172437** | **7705446** | **6577876** | **6455244** | **8813841** | **7977751** |

Each column represents a method of calculation from equation 1 - 4. The final two columns are the mid value of the four equations and the mean value of the four equations. Units are t of C.

Of the 8,813,841 t C lost across all estuaries pre-aquaculture to present 7,014,517 C t ± 27%, or 80%, can be attributed to direct displacement of mangrove forests by shrimp aquaculture (Figure 3) with approximately 34,500 ha of mangrove converted during the analysis period. Disregarding the federally protected Cayapas-Mataje estuary in which only 29.5% of mangrove loss is attributable to aquaculture, the CC loss due to this singular land use transition accounts for between 84% and 95% of all losses across all other estuaries. The most rapid losses of CC appear to occur at the initial period of shrimp aquaculture arrival within each estuary with greater than 72% of losses occurring between surveys one and three in each estuary.

Throughout the study period, particularly since 2000, there has been limited afforestation or reforestation within the estuaries resulted in CC increases (Figure 3). By 2011, a maximum potential of 1,709,079 ± 43% t C were added to the estuaries analyzed. However, this estimate is assuming complete maturity of the stands and is therefore an overestimation at the current juvenile stage of growth. The juvenile status was verified utilizing the date a stand first appears in a survey as well as being field verified in the respective estuaries. Almost all of this regrowth occurred in areas outside of the mangrove to shrimp conversion areas in the Chone and Muisné estuaries since 2000. At stand maturity (20 – 30 years) if undisturbed, this additional CC stock will offset 19% of the documented CC losses. The Chone and Muisné estuaries have experienced the largest reforestation/afforestation, with both having a maximum potential of 16% of their base level CC stocks replenished. The Cojimíes estuary has experienced the smallest level of recovery with, at most, a 5% addition from base CC levels.

**DISCUSSION**

When comparing our Ecuadorian CC data to IPCC GPG Tier 1 1 km$^2$ compliant carbon data based on GLC 2000 land cover classes, substantial differences occur at the estuarine level despite the country-wide value matching almost identically (Ruesch & Gibbs, 2008) (Table 2). Within Cayapas-Mataje, Chone and El Oro the IPCC results are in relatively close agreement with our CC findings (Table 2). The major differences between the two sets of results are in the neighboring estuaries of Muisné and Cojimíes. Within these estuaries the IPCC estuarine living carbon estimates are 2.16 and 4.08 times larger than our estimations. Part of this may be due to the differing scales of analysis causing forest vegetation on the estuarine edge to be included in the IPCC compliant data but this may not occur in our more resolute data. However, our results indicate this would result in a nominal amount of difference as most of the disagreement is in the central region of the estuaries.

Table 2 Mangrove CC levels vs. IPCC CC levels. MPD

| Estuary | IPCC 1km | Median | Difference | MPD |
|---|---|---|---|---|
| Cayapas - Mataje | 4253300 | 6243737 | 0.68 | 0.93 |
| Muisné | 417700 | 193936 | 2.15 | 1.70 |
| Cojimíes | 1921500 | 471478 | 4.08 | 3.21 |
| Chone | 277100 | 205416 | 1.35 | 1.06 |
| Isla Puna | 707700 | 1659324 | 0.43 | 0.58 |
| El Oro | 4380700 | 3239783 | 1.35 | 1.06 |
| **Total** | **11958000** | **12013673** | **1.00** | |

We report the median value of our findings closest to the year 2000 and the IPCC compliant findings for 2000. The MPD represents the Minimum Potential Difference when error bars are taken into account selecting the equation 1 - 4 that is closest to the IPCC measure.

The differences between our findings and the IPCC compliant data primarily occur due to the land cover classification schemes employed by the IPCC authors (GLC 2000 based on SPOT data) in which substantial broadleaved evergreen forests are shown to be present within Muisné and Cojimíes estuaries. The IPCC compliant methods do not contain a mangrove classification, so tropical evergreen broadleaf acts as the substitute. Our remote sensing surveys, field verification, and aerial imagery show little or no evidence of the existence of these forests as of 2000. Our data shows these forests, which are former mangrove areas, have been converted to aquaculture at earlier periods and hence the CC levels are substantially depleted. Within Chone estuary, the GLC 2000 herbaceous class is incorrectly shown to be dominant in the inner estuary. However, this only causes a slight over-estimation of living carbon as herbaceous cover has a far lower ecosystem carbon value than the misapplied evergreen forests depicted in the estuaries to the north. Within Chone our analysis again depicts these herbaceous regions as shrimp aquaculture. Within the Gulf of Guayaquil, on Isla Puna, the underestimation of carbon in the IPCC compliant data is also caused by classification differences, with our analysis finding some fringe mangrove in areas classified as water in the GLC 2000 dataset.

For the reasons above the IPCC compliant C grids within Ecuadorian estuaries should be treated with caution. Further research is needed to ascertain if this is a global problem or other nations follow the Ecuadorian pattern of large errors in GLC derived C measures at the estuarine level that combine into accurate C estimates at the national scale. For example, in Ecuador the GLC derived C underestimation and overestimation errors at the estuarine level essentially cancel themselves and sum to a national error rate of close to zero (Table 2). Despite summing to zero in the Ecuador example, the C errors are significantly large at the estuarine level that it cannot be assumed that all other nations over and under estimations will sum to zero as they do in the Ecuadorian example.

Our findings indicate that living carbon in Ecuador's mangrove forests has been substantially impacted by shrimp aquaculture expansion and that greater than 80% of mangrove carbon losses are a direct result of land use conversion to shrimp aquaculture. In estuaries where this conversion has occurred the IPCC compliant grids may be in substantial error at the estuarine level despite being in overall agreement nationally.

**NOTES**

We have provided a supplemental ESRI Map Package files. The first layer in the package represents CC losses in total and the second layer represents CC losses due to aquaculture. Data are present for each estuary, and the CC numbers provided are the aggregate of all survey periods within all the 1 ha grids. These files can be opened with the free ArcReader software

(http://www.esri.com/software/arcgis/arcreader) and will give the user full GIS access to our data. The results presented are measures of loss therefore any negative values equal CC gain.